\documentclass[fleqn,usenatbib,useAMS]{mnras}
\usepackage{multicol}        
\usepackage{bm}		
\usepackage{pdflscape}	

\usepackage{graphicx,epsf,epsfig}
\usepackage{inputenc} 
\usepackage{amsmath}
\usepackage{amsfonts}
\usepackage{mathrsfs}
\usepackage{amssymb}
\usepackage{amscd}
\usepackage{dcolumn}
\usepackage{bm}
\usepackage{natbib}
\usepackage{url}
\usepackage{xspace}

\usepackage[normalem]{ulem}
\usepackage{appendix}
\usepackage{algorithm}
\usepackage{algorithmic}

\usepackage{hyperref}
\usepackage{xcolor}

\def\be{\begin{equation}}
\def\ee{\end{equation}}
\def\bea{\begin{eqnarray}}
\def\eea{\end{eqnarray}}

\def\prl{Phys. Rev. Lett.}

\def\prd{Phys. Rev. D}

\def\mnras{Mon. Not. Roy. Astr. Soc.}
\def\apj{Astrophys. J.}
\def\apjl{Astrophys. J. Lett.}

\def\aap{Astron. Astrophys.}

\def\physrep{Phys. Rep.}
\def\jcap{J.C.A.P.}

\title{Numerical analysis of quasiperiodic oscillations in the Hartle-Thorne spacetime}

\author[K.~Boshkayev, et. al.]{K.~Boshkayev,$^{1,2,3}$\thanks{kuantay@mail.ru} 
T.~Konysbayev,$^{1,2}$\thanks{talgar\_777@mail.ru} Ye.~Kurmanov,$^{1,2}$\thanks{kurmanov.yergali@kaznu.kz} M.~Muccino,$^{2,4}$\thanks{marco.muccino@lnf.infn.it} and H.~ Quevedo,$^{2,5,6}$\thanks{quevedo@nucleares.unam.mx}\\
$^1$National Nanotechnology Laboratory of Open Type,  Almaty 050040, Kazakhstan.\\
$^2$Al-Farabi Kazakh National University, Al-Farabi ave. 71, 050040 Almaty , Kazakhstan.\\
$^3$Institute of Nuclear Physics, Ibragimova, 1, 050032 Almaty, Kazakhstan.\\
$^4$Universit\`a di Camerino, Via Madonna delle Carceri 9, 62032 Camerino, Italy\\
$^5$Instituto de Ciencias Nucleares, Universidad Nacional Aut\`onoma de M\`exico, Mexico.\\
$^6$Dipartimento di Fisica and Icra, Universit\`a di Roma “La Sapienza”, Roma, Italy.}

\begin{document}

\maketitle

\begin{abstract}
We numerically analyze quasiperiodic oscillations (QPOs) using a well-established spacetime model with neutron star sources. Within the framework of general relativity, we present expressions for the fundamental frequencies of test particles in the gravitational field of a slowly rotating and slightly deformed compact object defined by the Hartle-Thorne (HT) metric. Using the Relativistic Precession Model (RPM) formulated by Stella and Morinsk, we examine quasiperiodic oscillation data from eight neutron stars in low-mass X-ray binary systems. Employing Markov Chain Monte Carlo analyses with the Metropolis-Hastings algorithm, we estimate 1--$\sigma$ and 2--$\sigma$  error bars. Finally, we compare our results with predictions from the Schwarzschild, Lense-Thirring, and Kerr metrics, demonstrating that three of the eight sources can be well explained within the Hartle-Thorne model.
\end{abstract}


\section{Introduction}

Rotating neutron stars, which are frequently identified as pulsars, may exist in isolation or as part of binary systems. Within binary systems, these stellar objects can achieve elevated rotational frequencies because of the accretion of matter from their companion star. The accretion process increases both the gravitational mass and angular momentum of the neutron star, effectively boosting its rotational speed. The rotational frequencies of pulsars can reach up to several hundred hertz, with the most rapid pulsar recorded to date rotating at 716 Hz \citep{2006Sci...311.1901H}. This study predominantly focuses on neutron stars situated within low-mass X-ray binaries, where the neutron star is in association with a low-mass ordinary star. Observational data for some of these objects exhibit the characteristic phenomenon of twin high-frequency quasi-periodic oscillations (QPOs) \citep{Lewinvan2006}.

QPOs constitute a significant observational characteristic in high-energy astrophysical systems, notably within X-ray binaries and active galactic nuclei. These oscillations manifest as distinct peaks in the power spectral density of X-ray light curves, signifying periodic or semi-periodic modulations in the emitted radiation. Over time, QPOs demonstrate variations in both frequency and amplitude, thereby reflecting the intricate and dynamic nature of the underlying astrophysical processes. Numerous efforts have been made to elucidate the physical origin of QPOs through the application of straightforward orbital models based on the dynamics of matter in orbit around a central object (e.g.\, \citet{1998ApJ...492L..59S, 1999PhR...311..259W, 2001A&A...374L..19A, 2001AcPPB..32.3605K, 2001ASPC..234..213S, 2001ApJ...559L..25W, 2003A&A...404L..21A}). A comprehensive collection of references pertaining to neutron star QPO models, in addition to a direct comparative analysis of these models, is available in \citet{2010ApJ...714..748T, 2012ApJ...760..138T, 2016MNRAS.457L..19T}. A more advanced model that incorporated adjustments to the orbital dynamics owing to the finite thickness of the accretion disk was examined by \citet{2016MNRAS.457L..19T}. 
In our preceding publication \citep{2024MNRAS.531.3876B}, we analyzed QPO data from low-mass X-ray binary systems, employing the Lense-Thirring and Kerr metrics and the approximate Zipoy-Voorhees spacetime with quadrupole parameter. In the present paper, we consider the Hartle-Thorne metric, which describes the geometry around slowly rotating and slightly deformed astrophysical objects.

The exterior space-times surrounding rotating compact stars have been thoroughly investigated over the years. The seminal contributions of Hartle and Thorne involved the development of a slow-rotation approximation \citep{1967ApJ...150.1005H,1968ApJ...153..807H}, which elucidates the star's internal structure along with its surrounding vacuum space-time. This model is formulated as a perturbative extension of a spherically-symmetric non-rotating solution, with the perturbation considered up to the second order in the star's angular velocity $\Omega$. Within this approximation, the exterior space-time is comprehensively characterized by the gravitational mass $M$, the angular momentum $J$, and the quadrupole moment $Q$ of the rotating star.

This paper focuses on analyzing the circular and epicyclic trajectories of test particles in the vicinity of a compact object. In addition, we delve into the fundamental frequencies related to distinct metrics (models) and explore the QPOs observed in eight specific sources: Cir X-1, GX 5-1, GX 17+2, GX 340+0, Sco X-1, 4U1608-52, 4U1728-34, and 4U0614+091 \citep{2023PhRvD.108d4063B, 2022arXiv221210186B, 2024MNRAS.531.3876B, 2014GrCo...20..233B}. This study aims to investigate Hartle-Thorne solutions in scenarios where the source's deformation is small. Throughout this article, the signature of the line element is adopted as ($-$, $+$ ,$+$ ,$+$) and geometric units are used ($G=c=1$). However, expressions that have astrophysical relevance are presented in explicit physical units.

\label{intro}

\section{Quasiperiodic oscillations}\label{sezione2}
\subsection{The Hartle–Thorne metric}\label{sez2}
The Hartle--Thorne approach \citep{1967ApJ...150.1005H,1968ApJ...153..807H} provides a methodologically advantageous framework for modeling compact stars undergoing slow and rigid rotation. This approach has garnered significant attention and discourse among numerous scholars. Within this framework, deviations from spherical symmetry are treated as small perturbations. The line element for the Hartle-Thorne metric is \citep{
2024EPJP..139..273B, 2015ARep...59..441B} \\
\begin{align}\label{ht1}
ds^2=&-\left(1-\frac{2{M}}{r}\right)\nonumber\\
\times&\left[1+2k_1P_2(\cos\theta)+2\left(1-\frac{2{M}}{r}\right)^{-1} \frac{J^{2}}{r^{4}}(2\cos^2\theta-1)\right]dt^2 \nonumber\\  
+&\left(1-\frac{2{M}}{r}\right)^{-1}\left[1-2k_2P_2(\cos\theta)-2\left(1-\frac{2{M}}{r}\right)^{-1}\frac{J^{2}}{r^4}\right]dr^2\nonumber\\
+&r^2\Big[1-2k_3P_2(\cos\theta)\Big]\left(d\theta^2+\sin^2\theta d\phi^2\right) -\frac{4J}{r}\sin^2\theta dt d\phi\,,
\end{align}

\noindent with
\begin{eqnarray}\label{k123}
k_1&=&\frac{J^{2}}{{M}r^3}\left(1+\frac{{M}}{r}\right)+\frac{5}{8}\frac{Q-J^{2}/{M}}{{M}^3}Q_2^2\left(\frac{r}{{M}}-1\right),\nonumber\\
k_2&=&k_1-\frac{6J^{2}}{r^4},\nonumber\\
k_3&=&k_1+\frac{J^{2}}{r^4}+\frac{5}{4}\frac{Q-J^{2}/{M}}{{M}^2\left(r^2-2Mr\right)^{1/2}} Q_2^1\left(\frac{r}{M}-1\right)\,,\nonumber
\end{eqnarray}
and
\begin{eqnarray}
\label{ht2}
P_{2}(\cos\theta)&=&\frac{1}{2}(3\cos^{2}\theta-1),\nonumber\\
Q_{2}^{1}(x)&=&(x^{2}-1)^{1/2}\left[\frac{3x}{2}\ln\frac{x+1}{x-1}-\frac{3x^{2}-2}{x^{2}-1}\right],\nonumber\\
Q_{2}^{2}(x)&=&(x^{2}-1)\left[\frac{3}{2}\ln\frac{x+1}{x-1}-\frac{3x^{3}-5x}{(x^{2}-1)^2}\right],\nonumber
\end{eqnarray}
$x=r/M-1$, $P_{2}(x)$ denote the second Legendre polynomial of the first kind, $Q_l^m$ the associated Legendre polynomials of the second kind, while ${M}$, ${J}$, and ${Q}$ signify the total mass, angular momentum, and mass quadrupole moment of a rotating star, respectively. Additionally, $J\sim\Omega_{Star}$ and $Q\sim\Omega_{Star}^2$ are discussed, with $\Omega_{Star}$ representing the angular velocity of the central object, or star, which is the source of gravity.


\subsection{The frequencies of epicyclic motion}
The components of the metric tensor remain invariant under coordinate transformations of $t$ and $\phi$. This invariance indicates the presence of two constants of motion: the conserved specific energy per unit mass at infinity, denoted as $E$, and the conserved $z$ component of the specific angular momentum per unit mass at infinity, referred to as $L_z$. Consequently, this scenario permits the expression of the $t$- and $\phi$-components of the 4-velocity of a test particle as \citep{2012JCAP...09..014B}
\be
u^t = \frac{E g_{\phi\phi} + L_z g_{t\phi}}{
g_{t\phi}^2 - g_{tt} g_{\phi\phi}} \, , \qquad
u^\phi = - \frac{E g_{t\phi} + L_z g_{tt}}{
g_{t\phi}^2 - g_{tt} g_{\phi\phi}} \, .
\label{utuphiKerr}
\ee
From the normalization condition of the four-velocity, $g_{\mu\nu}u^\mu u^\nu = -1$,
one can write
\be
g_{rr}\dot{r}^2 + g_{\theta\theta}\dot{\theta}^2
= V_{\rm eff}(r,\theta,E,L_z) \, ,
\ee
where $\dot{r} = u^r = dr/d\lambda$, $\dot{\theta} = u^\theta = d\theta/d\lambda$,
$\lambda$ is the affine parameter along the geodesic curve, and the effective potential $V_{\rm eff}$ is given by
\be
V_{\rm eff} = \frac{E^2 g_{\phi\phi} + 2 E L_z g_{t\phi} + L^2_z
g_{tt}}{g_{t\phi}^2 - g_{tt} g_{\phi\phi}} - 1  \, .
\ee
Circular orbits in the equatorial plane are located at the zeros and the turning points of the effective potential: $\dot{r} = \dot{\theta} = 0$, which implies $V_{\rm eff} = 0$, and $\ddot{r} = \ddot{\theta} = 0$, requiring respectively $\partial_r V_{\rm eff} = 0$ and $\partial_\theta V_{\rm eff} = 0$. From these conditions, one can obtain the Keplerian angular velocity $\Omega_{\phi}$ for test particles\citep{2016EL....11630006B}:

\be
\Omega_{\phi} =
\frac{- \partial_r g_{t\phi}
\pm \sqrt{\left(\partial_r g_{t\phi}\right)^2
- \left(\partial_r g_{tt}\right) \left(\partial_r
g_{\phi\phi}\right)}}{\partial_r g_{\phi\phi}} \, .
\ee
Afterwards, the generic expression for the energy of test particles on circular orbit, 
\be\label{E}
E = - \frac{g_{tt} + g_{t\phi}\Omega_{\phi}}{\sqrt{-g_{tt} - 2g_{t\phi}\Omega_{\phi} - g_{\phi\phi}\Omega_{\phi}^2}} \, ,
\ee
leads to 
\be
   \label{eq:energy}
E=E_0\left[1\mp jF_{1}(r)+j^2 F_2(r)+q F_3(r)\right],  
\ee
for the Hartle-Thorne spacetime. Here, $E_{0}$ is the energy for the Schwarzschild space-time and the supporting functions, $F_{1;2;3}$, are given in Appendix \ref{Appendix}

Given the generic orbital angular momentum, 
\be\label{L}
L_z = \frac{g_{t\phi} + g_{\phi\phi}\Omega_{\phi}}{\sqrt{-g_{tt} - 2g_{t\phi}\Omega_{\phi} - g_{\phi\phi}\Omega_{\phi}^2}} \, ,
\ee
we compute it for the Hartle-Thorne metric,  obtaining
\be\label{eq:momentum}
   L_z= L_0\left[1\mp jH_1(r)+j^2 H_2(r)+q H_3(r)\right],   
\ee
where, analogously to the above case, $L_{0}$, say the  angular momentum  for the Schwarzschild metric and the supporting functions $H_{1;2;3}$ are reported in Appendix \ref{Appendix}

The sign ``$+$'' in $\Omega_{\phi}$ is for co-rotating (prograde) orbits and the sign ``$-$'' in $\Omega_{\phi}$ denotes counter-rotating (retrograde) orbits. The Keplerian frequency is simply $f_{\phi} =\Omega_{\phi}/2\pi$.
The orbits are stable under small perturbations if $\partial_r^2 V_{\rm eff} \le 0$ and $\partial_\theta^2 V_{eff} \le 0$. In the case of more general orbits, we have that  \citep{2005Ap&SS.300..143K}
\be\label{eq-or}
\Omega^2_r =- \frac{1}{2 g_{rr} (u^t)^2}
\frac{\partial^2 V_{\rm eff}}{\partial r^2} \, , 
\ee
\be
\Omega^2_\theta = - \frac{1}{2 g_{\theta\theta} (u^t)^2}
\frac{\partial^2 V_{\rm eff}}{\partial \theta^2} \, .
\label{eq-ot}
\ee
The radial epicyclic frequency is thus $f_r = \Omega_r/2\pi$ and the vertical one is $f_\theta = \Omega_\theta/2\pi$. For the Hartle-Thorne spacetime, it acquires the following form
\begin{equation}
 \label{eq:velocityphi}
\Omega_{\phi}=\Omega_{0\phi}\left[1\mp jA_1(r)+j^2 A_2(r)+qA_3(r)\right],
\end{equation}
where we use the following notations $j=J/M^2$, $q=Q/M^3$, and 
$\Omega_{0\phi}$ is the angular velocity for the Schwarzschild metric.
 
The radial epicyclic velocity $\Omega_r$ and the vertical epicyclic velocity $\Omega_{\theta}$ 
are given by
\begin{equation}
\Omega_r^2=\Omega_{0r}^2
[1+jB_{1}(r)-j^2B_{2}(r)-qB_{3}(r)],
\label{eq:velocityr}
\end{equation}
\begin{equation}
\Omega_\theta^2=\Omega_{0\theta}^2
[1- jC_{1}(r)+j^2C_{2}(r)+qC_{3}(r)].
\label{eq:velocitytheta}
\end{equation}
The functions  $\Omega_{0\phi},\Omega_{0r}^2,\Omega_{0\theta}^2, A_{1, 2, 3}, B_{1, 2, 3}$ and $C_{1, 2, 3}$ are detailed in Appendix \ref{Appendix}.

In the cases of $j\to 0$ and $q\to 0$, the Hartle-Thorne geometry reduces to the Schwarzschild geometry, and when $j^2\to 0$ and $q\to 0$, but $j\neq0$, one recovers the Lense-Thirring metric. Furthermore, the approximate Kerr geometry is recovered to second order in the dimensionless angular momentum $j=a/M$ when $q=j^2$ and particular coordinate transformations are used \citep{1968ApJ...153..807H,2016IJMPA..3141006B}. 

The RPM associates the lower QPO frequency $f_{L}$ with the frequency of the periastron precession, represented as $f_L=f_{\phi}-f_r$, and the upper QPO frequency $f_{U}$ with the Keplerian frequency, denoted as $f_{U}=f_{\phi}$.
Another physical quantity of great interest is the radius of the innermost stable circular orbit ($r_{ISCO}$). Using Eq. \eqref{E} or \eqref{L}, and imposing the condition $dE/dr=0$ or $dL/dr=0$ leads to an expression for  $r_{ISCO}$, which is given by
\citep{2016IJMPA..3141006B}
\begin{eqnarray}
 \label{eq:risco}
 \nonumber
r_{ISCO}&=6M\left[1\mp\frac{2}{3}\sqrt{\frac{2}{3}}j+\left(\frac{251647}{2592}-240\ln\frac{3}{2}\right)j^2\right. \\
&\left.+\left(-\frac{9325}{96}+240\ln\frac{3}{2}\right)q\right] \nonumber \\
&\approx 6M[1\mp0.5443j-0.2256j^2+0.1762q]\,.
\end{eqnarray}
with the $-$ and $+$ referring to co-rotating and counter-rotating motion, respectively. Only 
outside this orbit can quasi-circular geodesic motion be stable, giving rise to possibly 
observable quasi-periodic oscillatory effects.

Using the expression $f_U$, we find the radial coordinate, where QPOs originate, namely, 
\begin{equation}
 \label{eq:rfu}
\tilde{r}=\frac{M}{(2\kappa)^{2/3}}\left(1-j\frac{4\kappa}{3}-j^2D_1+qD_2\right),
\end{equation}
where  $\kappa=\pi Mf_U$ and $D_1$ and $D_2$ are given in Appendix \ref{Appendix} 

Eventually, this radial distance will allow one to calculate the inner and outer radii of the disk, using the maximum and minimum of $f_U$.

\section{Monte Carlo analysis}\label{sezione3}

We perform a MCMC analysis by means of the Metropolis-Hastings algorithm, searching for the best-fit parameters that maximize the log-likelihood.
\begin{equation}
\label{loglike}
    \ln \mathcal L = -\sum_{k=1}^{N}\left\{\dfrac{\left[f_{\rm L}^k-f_{\rm L}(p,f_{\rm U}^k)\right]^2}{2(\sigma f_{\rm L}^k)^2} + \ln(\sqrt{2\pi}\sigma f_{\rm L}^k)\right\}
\end{equation}
with $N$ data for each source, sampled as lower frequencies $f_{\rm L}^k$, attached errors $\sigma f_{\rm L}^k$, and upper frequencies $f_{\rm U}^k$.
The theoretically-computed frequencies $f_{\rm L}(p,f_{\rm U}^k)$ depend also on combinations of the parameters $p=\{M,\,j,\,q\}$, depending on the considered space-time.

We modify the \texttt{Wolfram Mathematica} code from \citet{2019PhRvD..99d3516A} and adapt it to the case of QPO data. We compute the log-likelihood from $\mathcal N\simeq 10^5$ total number of iterations covering the widest possible range of priors on the parameters of Schwarzschild and Hartle-Thorne metrics, allowing constraints that may contrast with the neutron-star interpretation:
\begin{equation}
M\in[0,5]\,{\rm M}_\odot \quad,\quad q\in[0,15] \quad,\quad j\in[-1,1]\,.
\end{equation}

\begin{table*}
\centering
\setlength{\tabcolsep}{1.em}
\renewcommand{\arraystretch}{1.2}
\begin{tabular}{llcccrrrrr}
\hline\hline
Source                                  & 
Metric                                  &
$M\,({\rm M}_\odot)$                    & 
$q$                                     & 
$j$                                     &
$\ln \mathcal L_{\rm max}$              &
AIC                                     &
BIC                                     &
$\Delta$AIC                             &
$\Delta$BIC                             \\
\hline
Cir X1                                  &
S                                       &
$2.22^{+0.03}_{-0.03}$                  & 
--                                      & 
--                                      & 
$-125.84$                               & 
$254$ & $254$ & $115$ & $114$           \\
                                        &
HT$_1$                                  & 
$3.88^{+0.07}_{-0.11}$                  & 
$0.05^{+0.16}_{-0.05}$                  & 
$0.59^{+0.04}_{-0.03}$                  & 
$-66.33$                                & 
$139$ & $140$ & $0$ & $0$               \\
                                        &
HT$_2$                                  & 
$3.31^{+0.16}_{-0.19}$                  & 
$0.01^{+0.05}_{-0.01}$                  & 
$0.55^{+0.05}_{-0.06}$                  & 
$-101.70$                               & 
$210$ & $211$ & $71$ & $71$             \\
\hline
GX 5-1                                  & 
S                                       &
$2.16^{+0.01}_{-0.01}$                  & 
--                                      & 
--                                      & 
$-200.33$                               & 
$403$ & $404$ & $186$ & $184$           \\
                                        &
HT$_1$                                  & 
$1.97^{+0.37}_{-0.21}$                  & 
$3.60^{+2.26}_{-0.90}$                  & 
$0.84^{+0.09}_{-0.12}$                  & 
$-105.06$                               & 
$216$ & $219$ & $0$ & $0$               \\
                                        &
HT$_2$                                  & 
$1.73^{+0.22}_{-0.17}$                  & 
$4.34^{+2.47}_{-1.51}$                  & 
$0.64^{+0.14}_{-0.13}$                  &  
$-105.30$                               & 
$217$ & $220$ & $1$ & $1$               \\
\hline
GX 17+2                                 &
S                                       &
$2.08^{+0.01}_{-0.01}$                  & 
$-$                                     & 
$-$                                     &    
$-1819.02$                              & 
$3640$ & $3640$ & $2188$ & $2187$       \\
                                        &
HT$_1$                                  & 
$1.54^{+0.01}_{-0.01}$                  & 
$0.01^{+0.04}_{-0.01}$                  & 
$<-0.79$                                & 
$-722.83$                               & 
$1452$ & $1453$ & $0$ & $0$             \\
                                        &
HT$_2$                                  &  
$1.23^{+0.02}_{-0.04}$                  & 
$4.03^{+0.74}_{-0.53}$                  & 
$<-0.79$                                &  
$-940.90$                               & 
$1888$ & $1889$ & $436$ & $436$         \\
\hline
GX 340+0                                & 
S                                       &
$2.10^{+0.01}_{-0.01}$                  & 
$-$                                     & 
$-$                                     &                                     
$-130.86$                               & 
$264$ & $264$ & $30$ & $29$             \\
                                        &
HT$_1$                                  & 
$2.05^{+0.28}_{-0.48}$                  & 
$0.84^{+0.64}_{-0.52}$                  & 
$0.15^{+0.21}_{-0.38}$                  & 
$-123.82$                               & 
$254$ & $255$ & $20$ & $20$             \\
                                        &
HT$_2$                                  &
$1.45^{+0.31}_{-0.12}$                  & 
$0.89^{+1.26}_{-0.88}$                  & 
$-0.56^{+0.34}_{-0.24}$                 &   
$-113.76$                               & 
$234$ & $235$ & $0$ & $0$               \\
\hline
Sco X1                                  &
S                                       &
$1.97^{+0.01}_{-0.01}$                  &
$-$                                     &
$-$                                     &
$-3887.17$                              &
$7776$ & $7778$ & $4354$ & $4350$       \\
                                        &
HT$_1$                                  & 
$1.38^{+0.01}_{-0.01}$                  & 
$<0.01$                                 & 
$<-0.79$                                & 
$-1708.08$                              & 
$3422$ & $3428$ & $0$ & $0$             \\
                                        &
HT$_2$                                  &  
$1.12^{+0.02}_{-0.01}$                  &
$3.61^{+0.22}_{-0.38}$                  &
$<-0.79$                                &
$-2471.71$                              &
$4949$ & $4955$ & $1527$ & $1527$       \\
\hline
4U1608–52                               &
S                                       &
$1.96^{+0.01}_{-0.01}$                  &
$-$                                     &
$-$                                     &
$-235.83$                               & 
$474$ & $474$ & $185$ & $184$            \\
                                        &
HT$_1$                                  & 
$1.38^{+0.02}_{-0.02}$                  & 
$0.04^{+0.16}_{-0.04}$                  & 
$<-0.79$                                & 
$-141.43$                               & 
$289$ & $291$ & $0$ & $0$               \\
                                        & 
HT$_2$                                  &
$1.21^{+0.06}_{-0.06}$                  &
$2.04^{+1.15}_{-0.90}$                  &
$<-0.74$                                &
$-184.97$                               &
$376$ & $378$ & $87$ & $87$           \\
\hline
4U1728–34                               &
S                                       &
$1.73^{+0.01}_{-0.01}$                  &
$-$                                     & 
$-$                                     &
$-212.61$                               & 
$427$ & $427$ & $136$ & $136$           \\
                                        &
HT$_1$                                  & 
$1.21^{+0.02}_{-0.02}$                  & 
$0.04^{+0.26}_{-0.04}$                  & 
$<-0.74$                                & 
$-142.37$                               & 
$291$ & $291$ & $0$ & $0$               \\
                                        &
HT$_2$                                  &
$1.06^{+0.06}_{-0.06}$                  & 
$1.94^{+1.26}_{-0.91}$                  & 
$<-0.73$                                & 
$-180.49$                               & 
$367$ & $367$ & $76$ & $76$             \\
\hline
4U0614+091                              &
S                                       &
$1.90^{+0.01}_{-0.01}$                  &
$-$                                     &
$-$                                     &
$-842.97$                               &
$1688$ & $1689$ & $622$ & $619$         \\
                                        &
HT$_1$                                  & 
$1.33^{+0.01}_{-0.01}$                  & 
$0.01^{+0.07}_{-0.01}$                  & 
$<-0.78$                                & 
$-530.20$                               & 
$1066$ & $1070$ & $0$ & $0$               \\
                                        &
HT$_2$                                      & 
$1.14^{+0.03}_{-0.04}$                  &
$2.37^{+0.76}_{-0.58}$                  &
$<-0.78$                                &
$-645.14$                               &
$1296$ & $1300$ & $230$ & $230$             \\
\hline
\end{tabular}
\caption{MCMC best-fit parameters, with the associated $1$--$\sigma$ error bars, for the Schwarzschild (S) and the Hartle-Thord (HT)  spacetimes with expansion orders in $J$ and $J^2$ or HT$_1$ ad HT$_2$, respectively. }
\label{tab:results} 
\end{table*}

To assess the best-fit model, we use the Aikake and the Bayesian Information Criterion, respectively AIC and BIC \cite{2007MNRAS.377L..74L}. Thus, considering our likelihood function from Eq.~\eqref{loglike}, we define
\begin{subequations}
\begin{align}
{\rm AIC}&=-2\ln \mathcal L_{\rm max}+2p\,,\\
{\rm BIC}&=-2\ln \mathcal L_{\rm max}+p\ln N\,,
\end{align}
\end{subequations}
where $\ln \mathcal L_{\rm max}$ is the maximum value of the log-likelihood, $p$ the number of estimated parameters in the model, and $N$ the number of the sample size. Recognizing the model with the lowest value of the AIC and BIC tests, say AIC$_{0}$ and BIC$_{0}$, as the fiducial (best-suited) model, the statistical evidence in support of the reference model is underlined by the difference $\Delta={\rm AIC(BIC)}-{\rm AIC_0(BIC_0)}$. 
Precisely, when comparing models, the evidence against the proposed model or, equivalently, in favor of the reference model can be naively summarized as follows:
\begin{itemize}
\item $\Delta{\rm AIC}$ and $\Delta{\rm BIC}\in[0,\,3]$, weak evidence;
\item $\Delta{\rm AIC}$ and $\Delta{\rm BIC}\in (3,\,6]$, mild evidence;
\item $\Delta{\rm AIC}$ and $\Delta{\rm BIC}>6$, strong evidence.
\end{itemize}

Besides the statistical selection criteria, each model has to pass also physical selection criteria, namely, it has to comply with NS theoretical expectations and match the observational constraints. Since we are interested in the mass, spin parameter and quadrupole parameter of neutron stars in accordance with the HT line element, from observation we have only the constraint on the minimum and maximum masses, which are around $\sim 1.174 M_\odot$ and $\sim 2.35 M_\odot$, respectively  \citep{2025PhRvL.134g1403M, 2022ApJ...934L..17R}. Owing to the intrinsic complexities of the problem, it is exceedingly difficult to directly measure the parameters $j$ and $q$, not to mention deriving observational constraints on them.

However, despite being model dependent, theoretically, one can always find constraints on $M$, $j$ and $q$.
\begin{itemize}
\item[-] \textit{Mass}. Depending on the EoS, NSs can have a maximum theoretical mass of 3.2 $M_\odot$ \citep{2002BASI...30..523S}.  
\item[-] \textit{Quadrupole}. The interval for possible values of $q$ for the most astrophysically interesting neutron-star models ranges from $q\thicksim 1$ for the most extreme objects close to maximal mass, and up to $q\thicksim 11$ for low mass objects \citep{2013MNRAS.433.1903U,2013ApJ...777...68B}. 
\item[-] \textit{Spin parameter}. A wide class of realistic equations of state (EoS) provides $j \sim 0.7$ \citep{2011ApJ...728...12L,2015PhRvD..92b3007C} and crust-less configurations enable $j\in(0.7,1]$ \citep{2016RAA....16...60Q}.
\end{itemize}

\begin{figure*}
{\hfill
\includegraphics[width=0.45\hsize,clip]{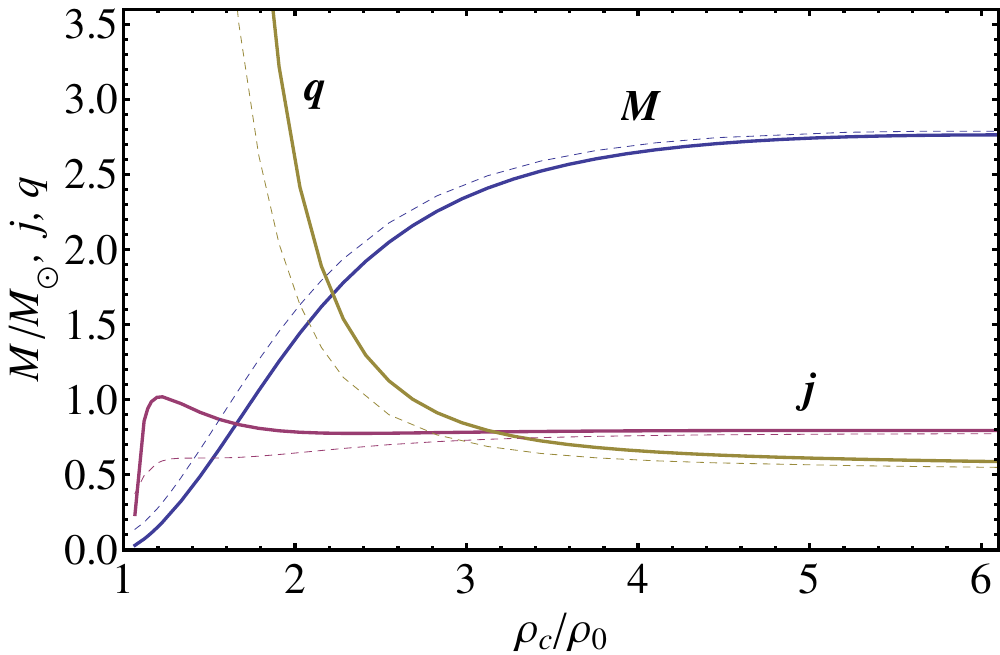}\hfill
\includegraphics[width=0.45\hsize,clip]{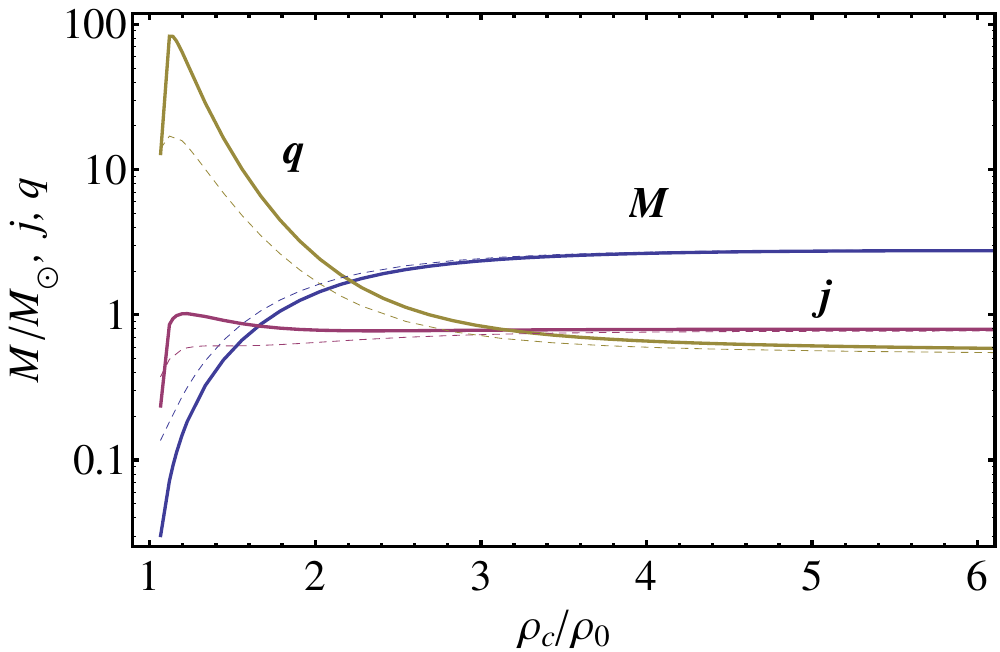}\hfill}
\caption{Total mass, spin parameter and quadrupole parameter as a function of the central density. Left linear scale. Right logarithmic scale.}
\label{fig:Mjqrho}
\end{figure*}

In Fig.~\ref{fig:Mjqrho}, for the maximally rotating NS model we show  $M$, $j$, and $q$ versus the central density $\rho_c$ of a NS normalized to the nuclear density $\rho_0$. Depending on the model, the parameters deviate slightly from each other. Nonetheless, for the increasing central density, the differences in the models decrease. It is obvious that for small masses, the values of $q$ can be large and vice versa, which is consistent  with other studies published in the literature \citep{2013MNRAS.433.1903U,2013ApJ...777...68B}. Furthermore, we will use these theoretical constraints to fit the QPO data. 

\section{Results and discussion}

\begin{figure*}
\centering
{\includegraphics[width=0.48\hsize,clip]{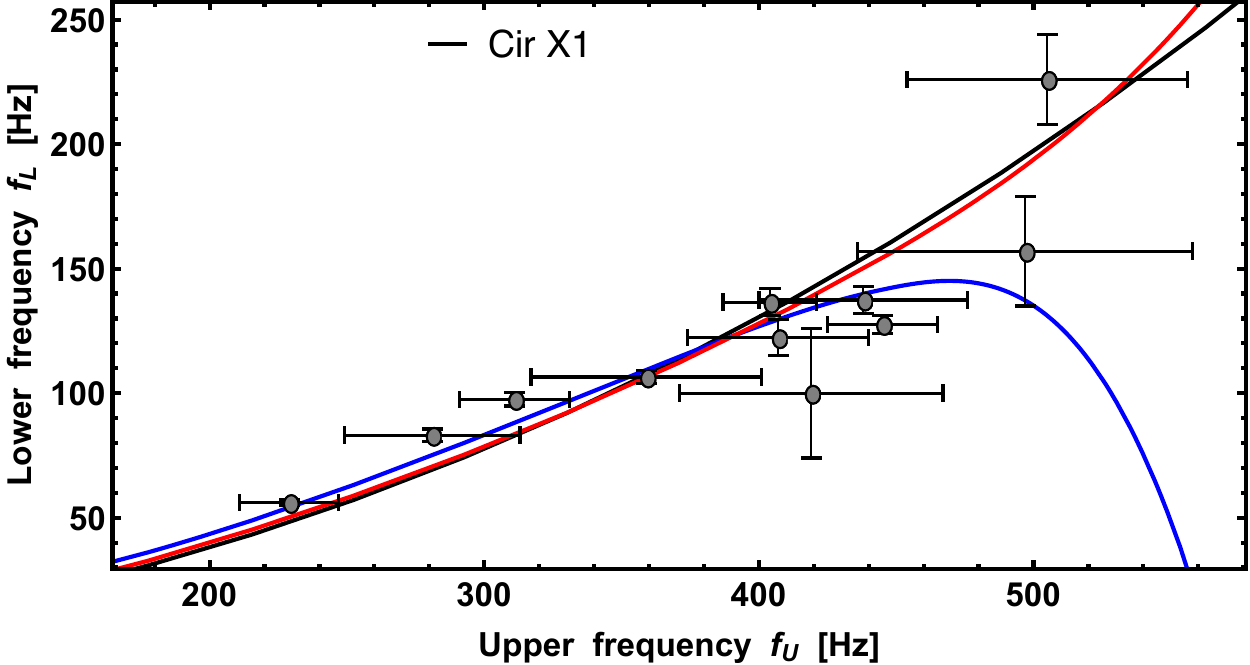}\hfill
\includegraphics[width=0.48\hsize,clip]{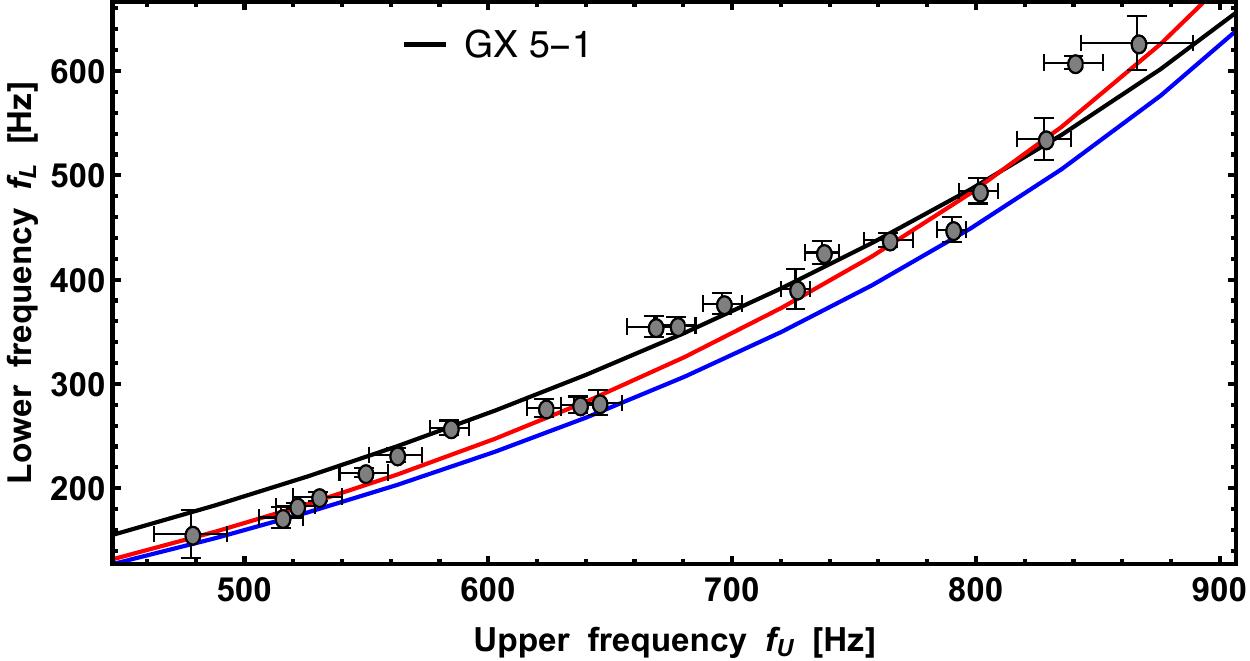}}
{\includegraphics[width=0.48\hsize,clip]{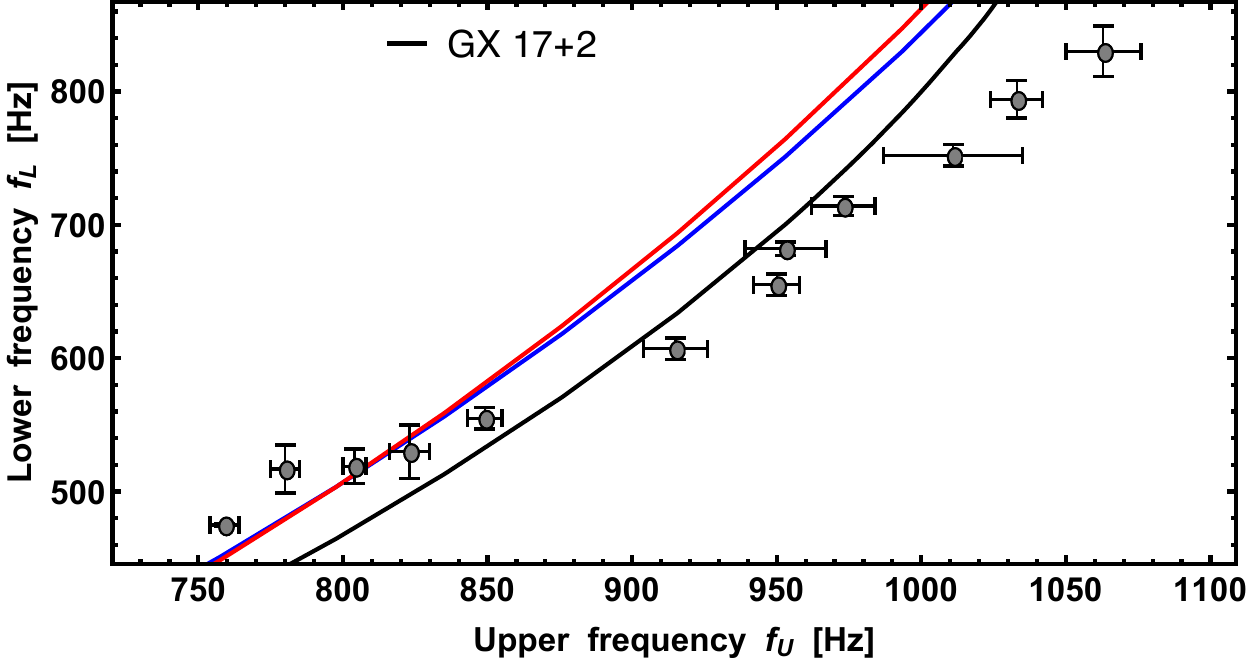}\hfill
\includegraphics[width=0.48\hsize,clip]{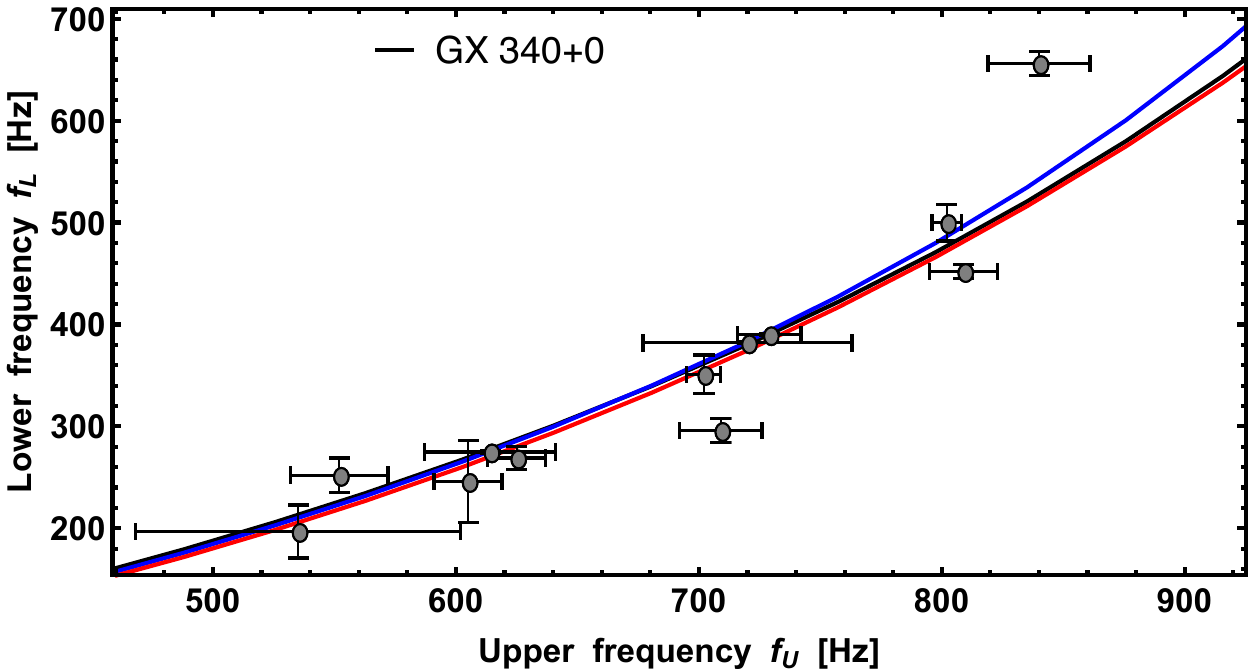}}
{\includegraphics[width=0.48\hsize,clip]{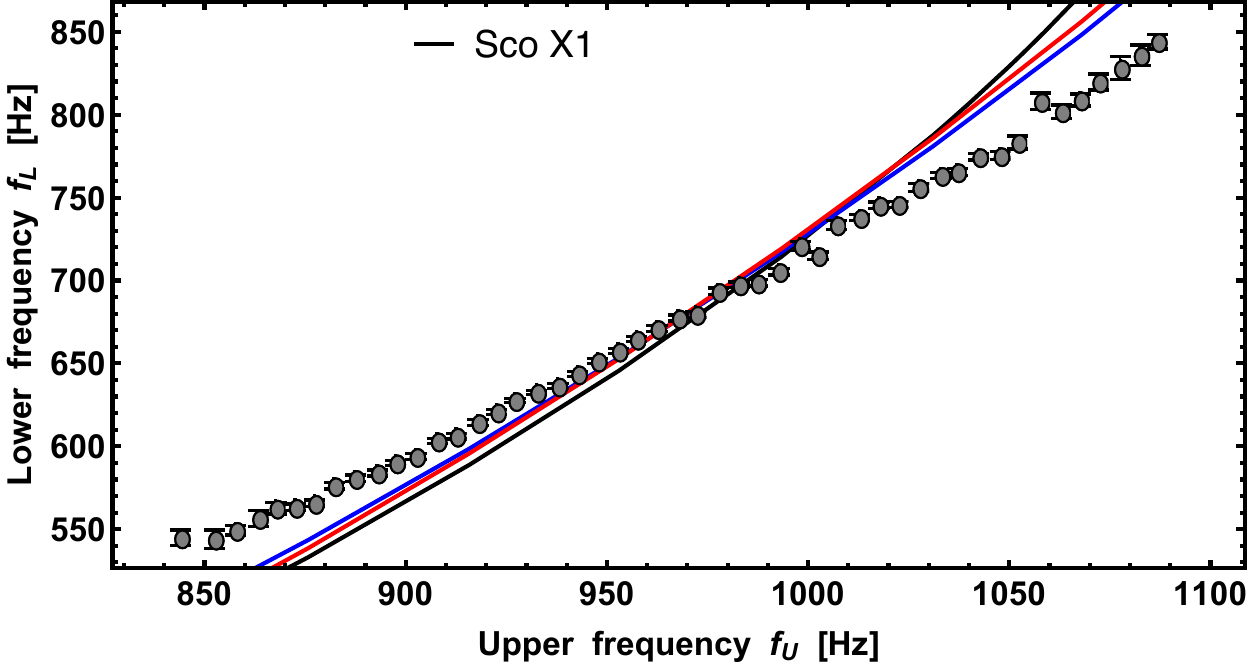}\hfill
\includegraphics[width=0.48\hsize,clip]{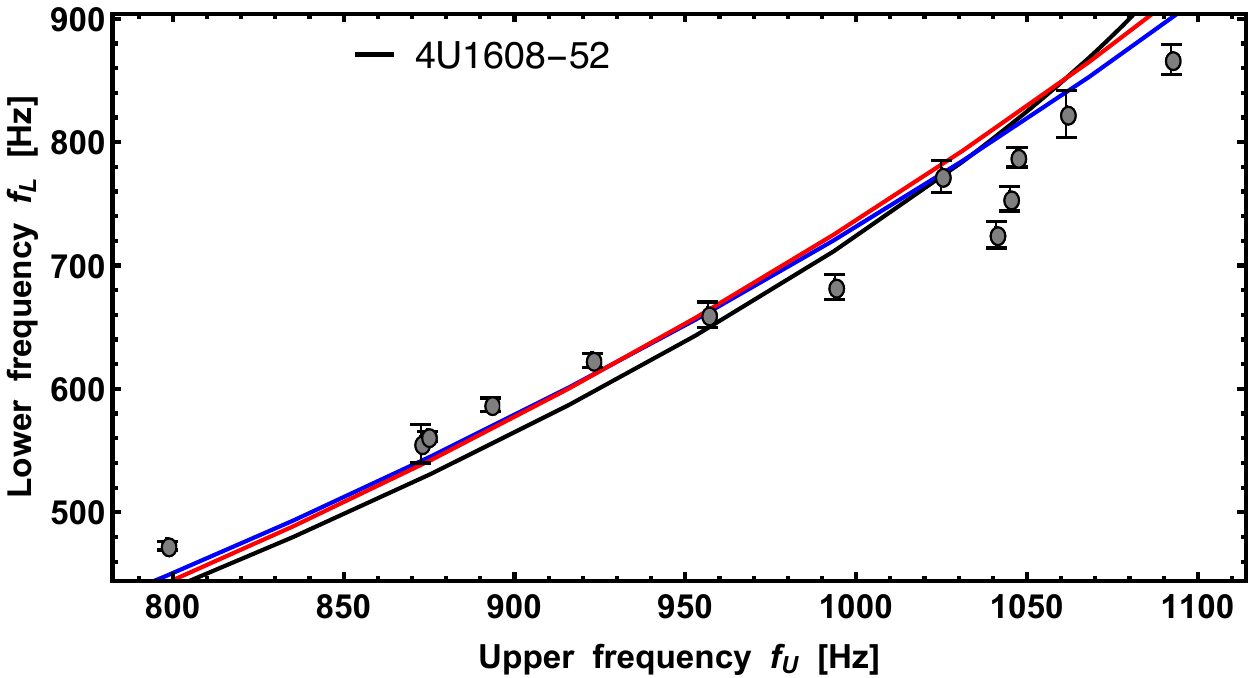}}
{\includegraphics[width=0.48\hsize,clip]{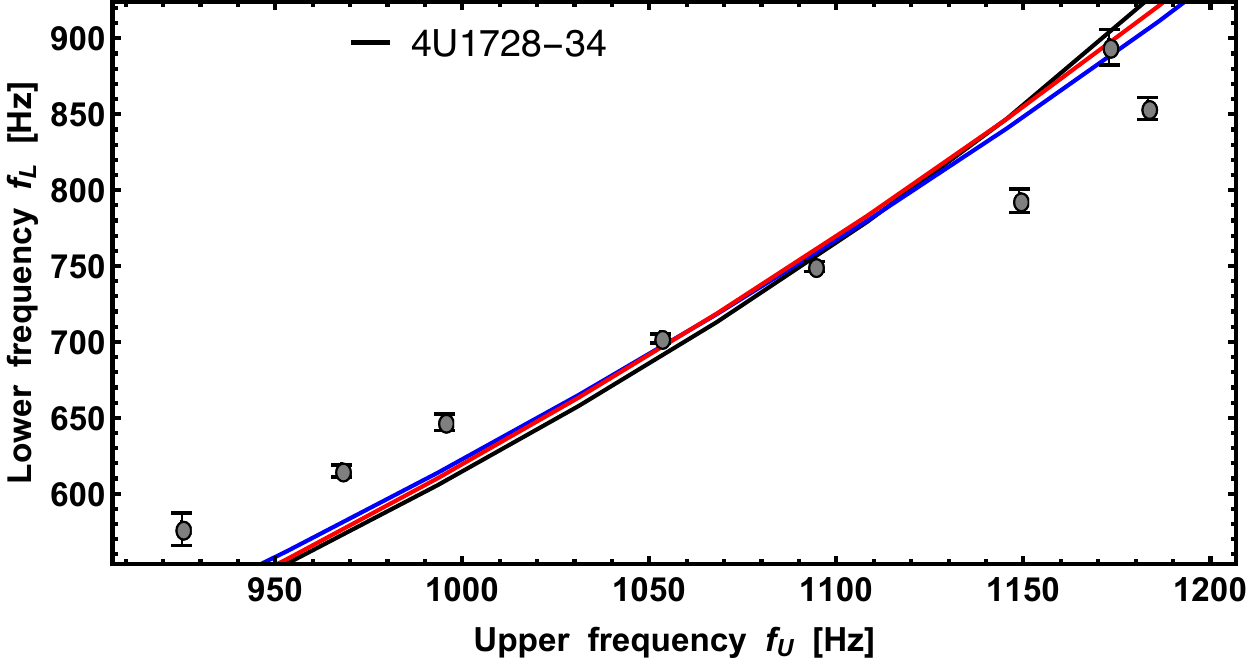}\hfill
\includegraphics[width=0.48\hsize,clip]{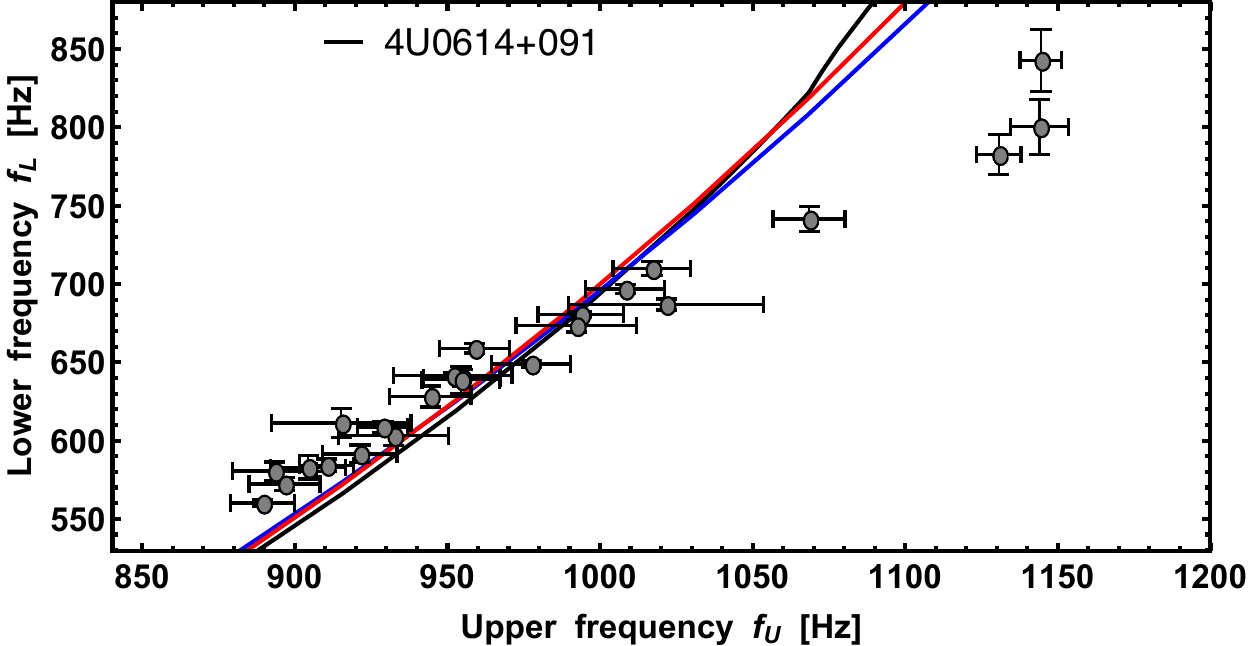}}
\caption{Plots of $f_{\rm L}$ vs $f_{\rm U}$ frequencies of the QPO data sets. The best-fit functions (solid lines) for the Schwarzschild (gray), HT$_1$ (blue), and HT$_2$ (red) space-times have been obtained from the best-fit parameters listed in Tab.~\ref{tab:results}.}
\label{fig:contours_HT2}
\end{figure*}

In our studies, we considered two cases of the Hartle-Thorne spacetime with expansion orders of $J$ and $J^2$ (HT$_1$ and HT$_2$, respectively). The results, summarized in Table~\ref{tab:results}, are analyzed statistically, observationally, and theoretically  below, for each source. Using Eqs. (\ref{eq:risco}) and (\ref{eq:rfu}), the numerical values of the ISCO and the radii of the inner and outer disks were calculated for each source (see Table~\ref{tab:isco}). For only three sources, namely Cir X1, GX 5-1 and GX 340+0, the ISCO of models HT$_1$,HT$_2$ provide physical results. The contour plots of the best-suited model parameters are shown in Figs.~\ref{fig:contours_HT}--\ref{fig:contours_HT2}.
In particular, Figs.~\ref{fig:contours_HT}--\ref{fig:contours_HT_II} display the normalized $1$--D loglikelihoods defined by the ratio $\ln \mathcal{L}/\ln \mathcal{L}_{\rm max}$.
\begin{table}
\centering
\setlength{\tabcolsep}{1.em}
\renewcommand{\arraystretch}{1.1}
\begin{tabular}{lcccc}
\hline
\hline
Source & Model & ISCO &  Inner &  Outer \\
       &       & (km) &  (km)  &  (km) \\
\hline
Cir X-1 & S & $19.67$ & $30.82$    &  $52.21 $  \\
$\,$ &  HT$_1$ & 23.64  & $38.04$ & $60.21$\\
$\,$ & HT$_2$ & 18.60 & $34.51$ & $59.13$\\
\hline
GX 5-1 & S & $19.14$    &  $21.31$ & $31.67$   \\
$\,$ &  HT$_1$ & $18.89$  & $22.36$ & $29.33$\\
$\,$ & HT$_2$ & $17.89$  & $19.67$ & $28.82$\\
\hline
GX 17+2 & S & $18.43$    &  $18.71$ & $22.57$   \\
$\,$ &  HT$_1$ &
19.61 & $16.49$ & $18.64$\\
$\,$ & HT$_2$ &
 21.81 & $16.55$ & $19.66$\\
\hline
GX 340+0 & S & $18.61$    &  $21.55$ & $29.11$   \\
$\,$ &  HT$_1$ & $17.70$  & $21.58$ & $25.82$\\
$\,$ & HT$_2$ & $17.87$  & $19.44$ & $26.05$\\
\hline
Sco X1 & S & $17.45$ &  $17.77$ & $21.03$   \\
$\,$ &  HT$_1$ & 17.55 & $15.39$ & $17.15$\\
$\,$ & HT$_2$ &   19.12 & $15.42$ & $18.02$\\
\hline
4U1608-52 & S & $17.37$ &  $17.68$ & $21.78$   \\
$\,$ &  HT$_1$ & 17.64 & $15.34$ & $18.26$\\
$\,$ & HT$_2$ &
17.59 & $15.92$ & $19.03$\\
\hline
4U1728-34 & S & $15.33$ &  $16.17$ & $18.94$   \\
$\,$ &  HT$_1$ & 15.46 & $14.02$ & $15.83$\\
$\,$ & HT$_2$ & 15.22 & $14.20$ & $16.50$\\
\hline
4U0614+091 & S & $16.83$ &  $19.49$ & $20.06$   \\
$\,$ &  HT$_1$ &16.94 & $16.95$ & $16.64$\\
$\,$ & HT$_2$ & 17.24& $16.96$ & $17.43$\\
\hline
\end{tabular}
\caption{Numerical values of ISCO and inner and outer disk radii for each source have been computed from best-fit results of Table~\ref{tab:results} for the Schwarzschild (S), HT$_1$ and HT$_2$ metrics.
}
\label{tab:isco}
\end{table}

\begin{figure*}
\centering
\includegraphics[width=0.98\hsize,clip]{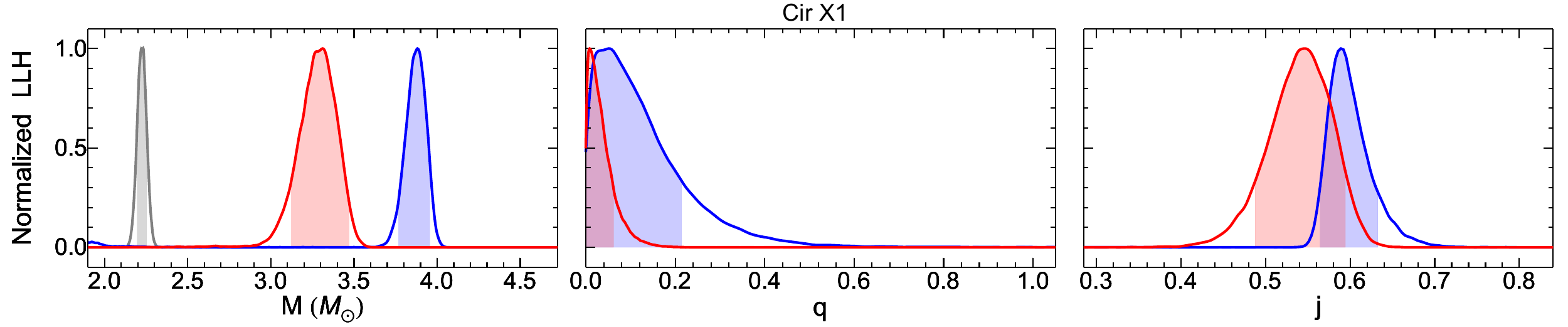}
\includegraphics[width=0.98\hsize,clip]{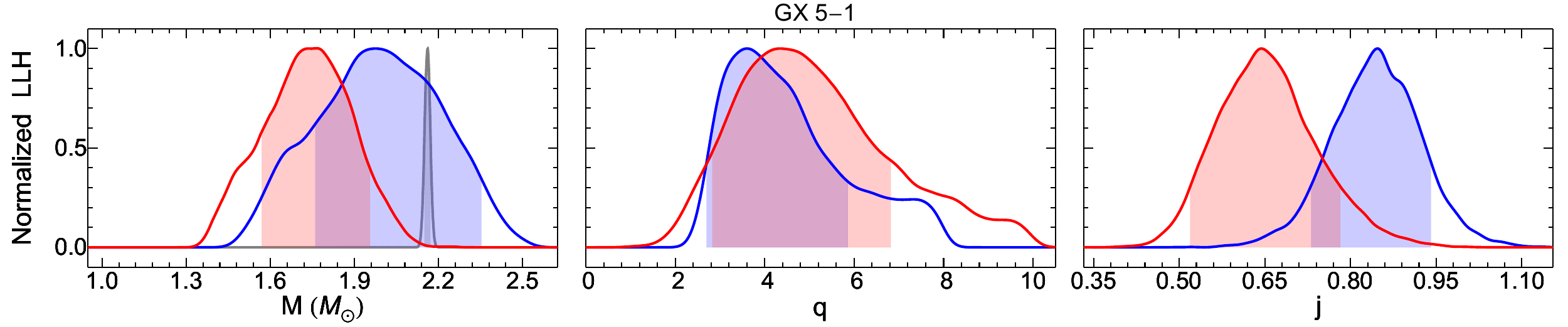}
\includegraphics[width=0.98\hsize,clip]{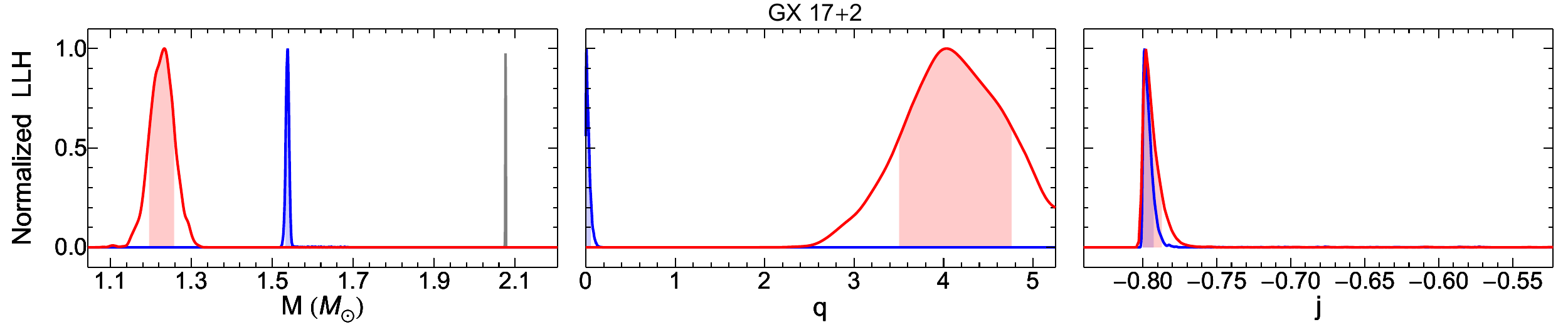}
\includegraphics[width=0.98\hsize,clip]{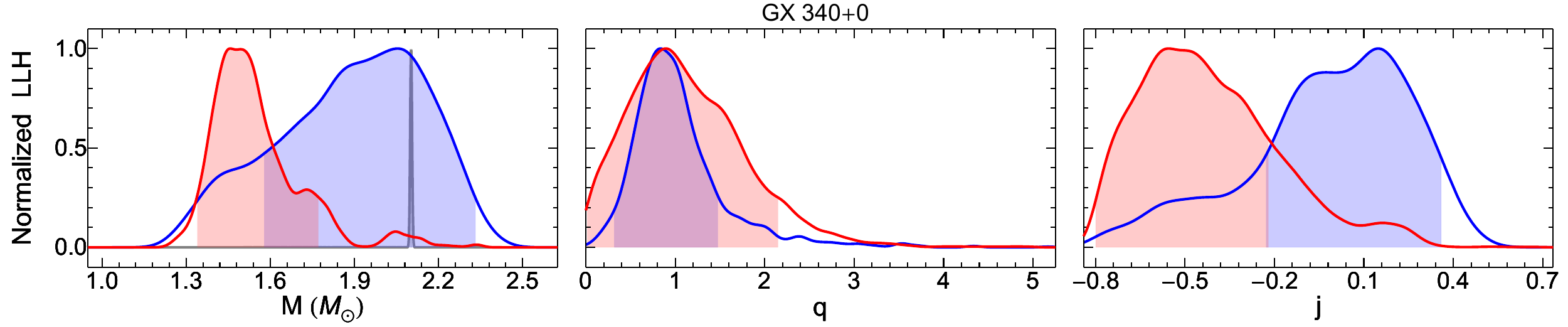}
\caption{Normalized $1$-D loglikelihood (LLH) functions (solid lines) and the $1$--$\sigma$ errors (shaded areas) of the Schwarzschild  (gray), HT$_1$ (blue), and HT$_2$ (red) spacetimes for each of the sources listed in Tab.~\ref{tab:results}.}
\label{fig:contours_HT}
\end{figure*}

\begin{figure*}
\centering
\includegraphics[width=0.98\hsize,clip]{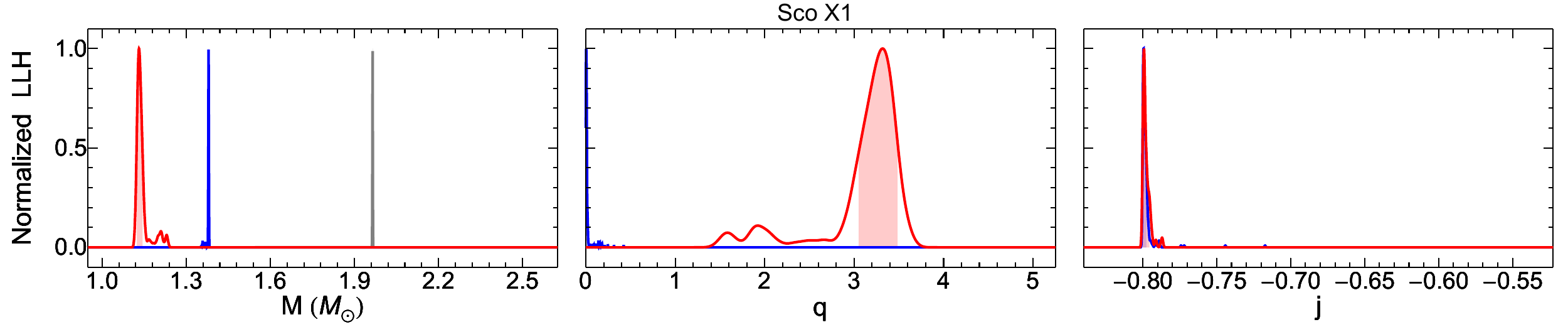}
\includegraphics[width=0.98\hsize,clip]{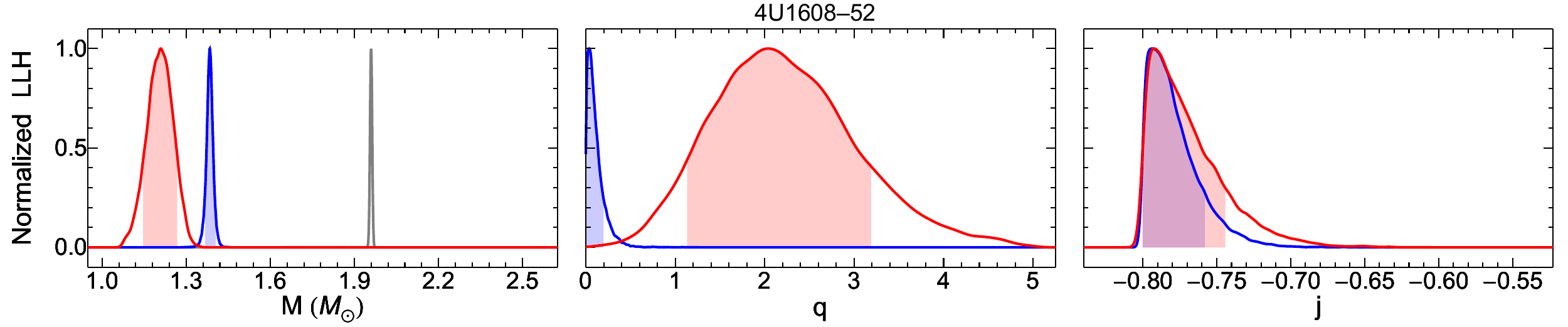}
\includegraphics[width=0.98\hsize,clip]{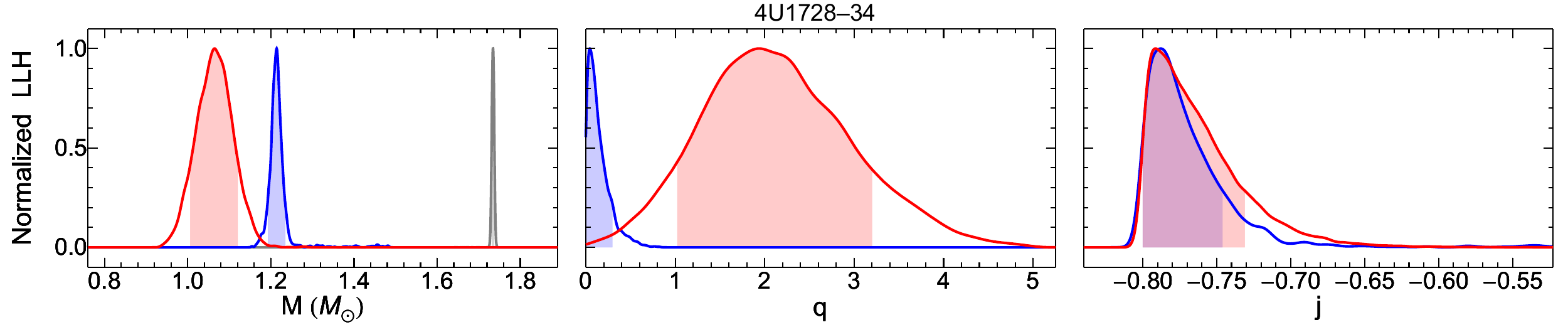}
\includegraphics[width=0.98\hsize,clip]{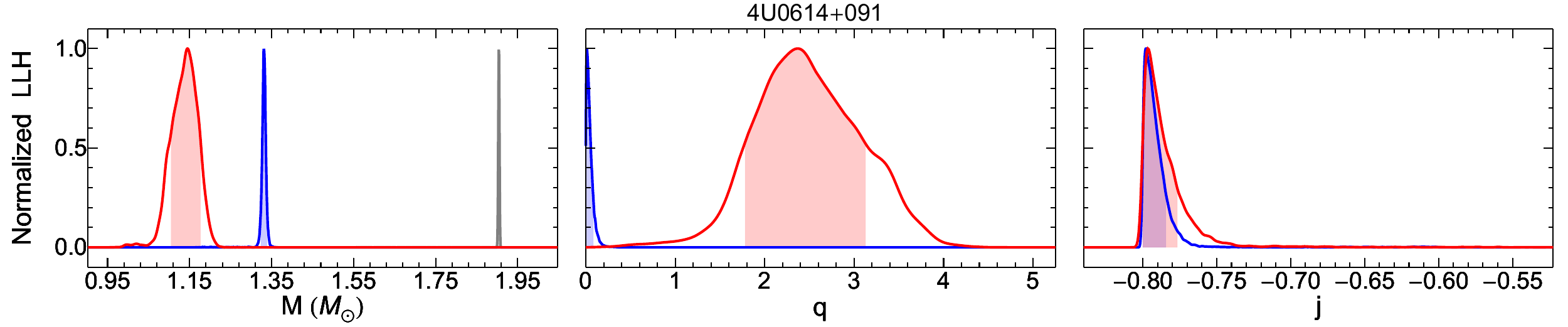}
\caption{Continued Fig.~\ref{fig:contours_HT}.}
\label{fig:contours_HT_II}
\end{figure*}

\begin{itemize}
\item[-] {\bf Cir~X-1} \cite{2006ApJ...653.1435B}.
As seen from Table~\ref{tab:results}, the HT$_{1}$ space-time is favored in comparison to the HT$_{2}$ space-time and Schwarzschild models. These solutions yield appropriate physical values for the ISCO, as detailed in Table~\ref{tab:isco}.
Nevertheless, in the cases of HT$_{1}$ and HT$_{2}$, the resulting masses $M\gtrsim 3$~M$_\odot$ appear to be incompatible with the 
neutron-star interpretation as advocated in Ref.~\cite{2010ApJ...714..748T}.
However, when examining the absolute maximum boundary of $M_{\rm up}=6.1$~M$_\odot$ \cite{2002BASI...30..523S}, the analyses with HT$_{1}$ and HT$_{2}$ indicate a mass lower than $M_{\rm up}$.
Consequently, we deduce that from statistical, theoretical, and experimental standpoints, HT$_{1}$ emerges as the preferred solution.
\item[-] {\bf GX~5-1} \cite{1998ApJ...504L..35W,2002MNRAS.333..665J}.
An examination of Tables~\ref{tab:results}--\ref{tab:isco} concerning the source GX~5-1 reveals that all models offer satisfactory fits to the data and reasonable physical values for the ISCO. Statistically, the HT$_{1}$ metric, with its parameters, emerges as the benchmark model that yields a well-constrained mass, aligning with existing observations of 
neutron-star masses. Consequently, we identify it as the best-fit model. However, for the HT$_{1}$ metric, the physical magnitude delineated in source $j=0.84$ is large for neutron stars as indicated in Ref. \citet{2011ApJ...728...12L}. Moreover, the mass specified in $M=1.97 M_\odot$ is consistent with the majority of neutron-star models, and the value of the spin parameter $j=0.84$ can be rationalized by the absence of the neutron-star crust, as noted in Ref. \citep{1999ApJ...512..282L}. 

{\bf GX~340+0} \cite{2000ApJ...537..374J}.
From Table~\ref{tab:results}, the optimal fit is indicated by the HT$_{2}$ metric, which is moderately favored over the HT$_{1}$ metric and aligns with observational data \cite{2013Sci...340..448A}. Furthermore, according to Table~\ref{tab:isco}, all metrics present valid physical ISCO values. The values of $q$ for all models are in good agreement with Refs. \citep{2013MNRAS.433.1903U,2013ApJ...777...68B} and Fig.~\ref{fig:Mjqrho}. With regard to the paramete $j$,  HT$_{1}$ is physically more favorable than HT$_{2}$.  Accordingly, it can be inferred that from a statistical perspective, HT$_{2}$ emerges as the optimal solution. Because the values of $BIC(AIC)=235(234)$ are minimal for this model (See Tab.~\ref{tab:results}). 

\item[-] {\bf 4U1608-52} \cite{1998ApJ...505L..23M}, {\bf 4U1728-34}  \cite{1999ApJ...517L..51M}, {\bf 4U0614+091} \cite{1997ApJ...486L..47F}.
The results obtained for the three sources   mentioned above are well described by the HT$_{1}$ metric rather than HT$_{2}$. However, the parameter $j$ is large ($j<-0.74$) for neutron stars as showed in \citet{2011ApJ...728...12L}, and the negative sign indicates that the accretion disk rotates counter to the direction of the central object, which is theoretically feasible. However, the mass range values of $1.06 M_\odot$ to $1.38 M_\odot$ and the quadrupole values from $0.01$ to $2.37$ for the HT$_{1}$ and HT$_{2}$ models align well with those reported in the existing literature. Moreover, the ISCO values are not physical for these two models (see Tab.~\ref{tab:isco}).

\item[-] {\bf Sco~X1} \cite{2000MNRAS.318..938M}, {\bf GX~17+2} \cite{2002ApJ...568..878H}.
The HT$_{1}$ metric provides the optimal statistical fit for these sources. Similar to the class 4U sources, the $j$ parameter is notably large ($j<-0.79$). The solutions HT$_{1}$ and HT$_{1}$ provide ``good'' masses for neutron stars, which are close to the standard mass $M=1.4M_\odot$. However, for these sources, the values of the ISCO are not physically meaningful.

\end{itemize}

\section{Conclusion}\label{sezione5}
In this study, we concentrated on eight NSs sources, analyzing the observed frequency data via three QPO models, within the context of the relativistic precession. We considered here two models of the Hartle-Thorne metric with expansion orders $J$ and $J^2$ (HT$_1$ and HT$_2$, respectively). In addition to these two models, for comparison, we also considered the standard Schwarzschild solution. Using the MCMC fitting procedure based on the Metropolis-Hastings algorithm, the Hartle-Thorne solutions were analyzed. Furthermore, the results of our analysis were evaluated and contrasted using the AIC and BIC statistical tests. Therefore, for each QPO data set, we calculated the best fit values, in particular, for the mass, spin, quadrupole, and ISCO parameters.

Statistically, the HT$_1$ metric gave the best fit for seven of the eight sources. Only for GX 340+0,  the HT$_2$ model showed the best result. For seven sources, the HT$_1$ mass fits the range $M \sim 1.21-2.05 M_\odot$, which are realistic values according to Ref. \citep{2002BASI...30..523S}. Furthermore, for GX 17 + 2, Sco X1, 4U1608-5, 4U1728-34, and 4U0614 + 091, the spin parameter $j$ showed the highest values ($j<-0.74$), which in turn indicates counter-rotating orbits. Moreover, for Cir X-1, GX 5-1 and GX 340+0, the HT$_1$ model gives a reasonable value for the spin parameter with $j \sim 0.15-0.84$. The values of the quadrupole $q$ for all sources range from 0.01 to 4.34 for the HT$_1$ and HT$_2$ models, which are in good agreement with the data reported in the existing literature \citep{2013MNRAS.433.1903U,2013ApJ...777...68B}. Regarding the ISCO value for the HT$_1$ and HT$_2$ models, the five sources GX 17+2, Sco X1, 4U1608-5, 4U1728-34, and 4U0614+091 are not physical. The Schwarzschild metric persistently demonstrated the worst fit, which can likely be ascribed to its incorporation of only one single free parameter in comparison to other metrics. However, from both physical and observational viewpoints, the Schwarzschild metric produces viable and reliable mass estimates.

 For sources with unconstrained values of $j$ (see Table~\ref{tab:results}), we have also explored priors far beyond those allowed for NSs, i.e., $j\in[-1000,1000]$. However, the results were not encouraging, since the MCMC analyses did not converge to well-determined and within-the-priors values of $j$ and, even worse, the masses dropped to unrealistic values $M<0.1$--$0.2$~M$_\odot$.

Unfortunately, direct measurements of the masses of compact components in LMXBs remain unattainable. This absence poses substantial challenges in the evaluation of theoretical models of NSs. The observational constraints, which have the potential to diverge from the theoretical predictions, are predominantly derived from a limited array of sources \citep{2011PrPNP..66..674T}. As a result, it is not consistently feasible to make direct comparisons between our findings and observational data.

\section*{Acknowledgements}
TK acknowledges grant No. AP19174979, YeK acknowledges grant No. AP19575366, and KB and MM acknowledge grant No. BR21881941, all from the Science Committee of the Ministry of Science and Higher Education of the Republic of Kazakhstan. The work of HQ was supported by PAPIIT-DGAPA-UNAM, grant No. 108225 and Conahcyt, grant No. CBF-2025-I-243.  

\section*{Data availability}
Data for quasiperiodic oscillations have been sourced from publicly accessible publications:
\cite{2006ApJ...653.1435B}, \href{https://doi.org/10.1086/508934}{DOI:10.1086/508934},
\cite{1998ApJ...504L..35W}, \href{
https://doi.org/10.1086/311564}{DOI:10.1086/311564}, \cite{2002MNRAS.333..665J}, \href{
https://doi.org/10.1046/j.1365-8711.2002.05442.x}{DOI:10.1046/j.1365-8711.2002.05442.x},
\cite{2002ApJ...568..878H}, \href{
https://doi.org/10.1086/339057}{DOI:10.1086/339057},
\cite{2000ApJ...537..374J}, \href{https://doi.org/10.1086/309029}{DOI:10.1086/309029},
\cite{2000MNRAS.318..938M}, \href{
https://doi.org/10.1046/j.1365-8711.2000.03788.x}{DOI:10.1046/j.1365-8711.2000.03788.x},
\cite{1998ApJ...505L..23M}, \href{https://doi.org/10.1086/311600}{DOI:10.1086/311600},
\cite{1999ApJ...517L..51M}, \href{
https://doi.org/10.1086/312025}{DOI:10.1086/312025},
\cite{1997ApJ...486L..47F},  \href{https://doi.org/10.1086/310827}{DOI:10.1086/310827}.
In addition, for some sources the data points were kindly provided by Professor Mariano M{\'e}ndez via private communication.

\appendix
\section{Circular orbits in the Hartle-Thorne space-time} 
\label{Appendix}

This section provides the expressions involved in the formulas for angular velocity, angular momentum, and energy.

The orbital angular momentum for circular particle orbits is given
by Eq. (\ref{eq:momentum}), where

\begin{eqnarray}  
 L_{0}(r)&=&r\sqrt{\frac{M}{r-3M}}, \label{eq:L0}
      \end{eqnarray}
and the functions $H_{1;2;3}$ can be expressed as
      \begin{eqnarray} 
            H_{1}(r)&=&\frac{3M^{3/2}(r-2M)}{r^{3/2}(3M-r)},  \label{eq:H1} \\
    H_{2}(r)&=&\left[16M^2r^4(r-3M)^{2}\right]^{-1}\Big[144M^8  \label{eq:H2} \nonumber \\
            &-&144M^7r+20M^6r^2-98M^5r^3 \nonumber \\
            &+&147M^4r^4+205M^3r^5-260M^2r^6\nonumber \\
            &+&105Mr^7-15r^8\Big]+H(r) , \\
    H_{3}(r)&=&5\left[16M^2r(3M-r)\right]^{-1}\Big(6M^4-7M^3r  \label{eq:H3}\nonumber\\
            &-&16M^2r^2+12Mr^3-3r^4\Big)-H(r) ,\\
        H(r)&=&15\left[32M^3(3M-r)\right]^{-1}\Big(6M^4+2M^3r  \label{eq:H}\nonumber\\
            &-&9M^2r^2+5Mr^3-r^4\Big) \ln\left(\frac{r}{r-2M}\right).
\end{eqnarray}
The energy of test particles on circular orbits is given by Eq. (\ref{eq:energy}) with 
\begin{eqnarray}
 \label{eq:E0}
E_0(r)=\frac{r-2M}{\sqrt{r(r-3M)}}, 
\end{eqnarray}
and the functions $F_{1;2;3}$ are
\begin{eqnarray}
 \label{eq:F1}
F_{1}(r)=\frac{M^{5/2}r^{-1/2}}{(r-2M)(r-3M)}, 
\end{eqnarray}
\begin{eqnarray}
F_2(r)&=&\left[16Mr^4(2M-r)(r-3M)^{2}\right]^{-1} \label{eq:F2}\nonumber\nonumber\\ 
&\times&\Big(-144M^8+144M^7r+28M^6r^2\nonumber\\ 
&+&58M^5r^3+176M^4r^4-685M^3r^5\nonumber\\ 
&+&610M^2r^6-225Mr^7+30r^8\Big)-F(r),   \\
F_{3}(r)&=&-5\left[16Mr(r-2M)(r-3M)\right]^{-1} \label{eq:F3} \nonumber\\
&\times&\Big(6M^4+14M^3r-41M^2r^2\nonumber\\
&+&27Mr^3-6r^4\Big)+F(r),\\
F(r)&=&\frac{15r(8M^2-7Mr+2r^2)}{32M^2(3M-r)} \ln\left(\frac{r}{r-2M}\right) . \quad  \label{eq:F}
\end{eqnarray}

The angular velocity for corotating/counterrotating circular particle orbits is given by Eq. (\ref{eq:velocityphi}), where  
\begin{eqnarray}
   \Omega_{0\phi}(r)&=&\frac{M^{1/2}}{r^{3/2}},
   \end{eqnarray}
and the functions $A_{1, 2, 3}$ are 
   \begin{eqnarray}
    A_1(r)&=&\frac{M^{3/2}}{r^{3/2}},\\
    A_2(r)&=&\left[16M^2r^4(r-2M)\right]^{-1}\Big(48M^7-80M^6r \nonumber\\
          &+&4M^5r^2-18M^4r^3+40M^3r^4+10M^2r^5\nonumber \\
           &+&15Mr^6-15r^7\Big)+A(r) \\
     A_3(r)&=&5\left[16M^2r(r-2M)\right]^{-1}\Big(6M^4-8M^3r \nonumber\\
         &-&2M^2r^2-3Mr^3+3r^4\Big)-A(r), \\
     A(r)&=&\frac{15(r^3-2M^3)}{32M^3}\ln\left(\frac{r}{r-2M}\right).
\end{eqnarray}
 
The radial epicyclic velocity for circular particle orbits is given
by Eq. (\ref{eq:velocityr}) and
\begin{eqnarray}
    \Omega_{0r}^2(r)&=&\frac{M^3(r-6M)}{r^{4}},\\
    B_{1}(r)&=&\frac{6M^{3/2}(r+2M)}{r^{3/2}(r-6M)},\\
    B_{2}(r)&=&\left[8M^2r^4(r-2M)(r-6M)\right]^{-1}\Big(384M^8-720M^7r\nonumber\\&-&112M^6r^2-76M^5r^3-138M^4r^4-130M^3r^5\nonumber\\&+&635M^2r^6-375Mr^7+60r^8\Big)+B(r),\\
    B_{3}(r)&=&5\left[8M^2r(r-2M)(r-6M)\right]^{-1}\Big(48M^5+30M^4r\nonumber\\&+&26M^3r^2-127M^2r^3+75Mr^4-12r^5\Big)-B(r)\\
    B(r)&=&\frac{15r(r-2M)(2M^2+13Mr-4r^2)}{16M^3(r-6M)}\ln\left(\frac{r}{r-2M}\right).
\nonumber
\end{eqnarray}

The vertical epicyclic velocity for circular particle orbits is given
by Eq. (\ref{eq:velocitytheta}) and
\begin{eqnarray}
\Omega_{0\theta}^2(r)&=&\frac{M^3}{r^{3}},\\
    C_{1}(r)&=&\frac{6M^{3/2}}{r^{3/2}},\\
    C_{2}(r)&=&\left[8M^2r^4(r-2M\right]^{-1}\Big(48M^7-224M^6r\nonumber\\&+&28M^5r^2+6M^4r^3-170M^3r^4+295M^2r^5\nonumber\\&-&165Mr^6+30r^7\Big)+C(r)\\
    C_{3}(r)&=&\frac{5(6M^4+34M^3r-59M^2r^2+33Mr^3-6r^4)}{8M^2r(r-2M)}+C(r),\\
    C(r)&=&\frac{15(2r-M)(r-2M)^2}{16M^3}\ln\left(\frac{r}{r-2M}\right).
\end{eqnarray}

The radial coordinate is given by Eq. (\ref{eq:rfu}) and
\begin{eqnarray}
\label{eq:rj2}
D_1&=& \frac{28 \kappa ^2}{9}+\frac{1}{48\kappa^{4/3}\left(4\kappa ^{2/3}-2^{1/3}\right)}\nonumber\\ &\times&
\Bigg[768(2\kappa^7)^{2/3}+32(2\kappa^{10})^{1/3}-72(2\kappa^{4})^{2/3}+20(2\kappa^{4})^{1/3}\nonumber\\
&+&15(2\kappa)^{2/3}-1280\kappa^4+160\kappa^2-15\Bigg]+D,\\
D_2&=& D-\frac{5\Big(3-3(2\kappa)^{2/3}-32\kappa^2+24(2\kappa^{4})^{2/3}\Big)}{48\kappa^{4/3}\left(4\kappa ^{2/3}-2^{1/3}\right)},\\
    D&=&\frac{5\left(8\kappa^2-1\right)\ln \left(\frac{2}{\left(2^{1/3}-4 \kappa^{2/3}\right)^3}\right)}{192\kappa^2}.
   \end{eqnarray}

\bibliographystyle{mnras}
\bibliography{0refs} 
\end{document}